\documentclass[]{spie}  %>>> use for US letter paper
\usepackage{amsmath,amsfonts,amssymb}
\usepackage{graphicx}
\usepackage[colorlinks=true, allcolors=blue]{hyperref}

\title{Seeing Beyond Cancer: Multi-Institutional Validation of Object Localization and 3D Semantic Segmentation using Deep Learning for Breast MRI}

\author[a]{Arda Pekis}
\author[a]{Vignesh Kannan}
\author[a]{Evandros Kaklamanos}
\author[a]{Anu Antony}
\author[a]{Snehal Patel}
\author[a]{Tyler Earnest}
\affil[a]{SimBioSys, Inc., Chicago, IL, USA}

\authorinfo{Further author information: Send correspondence to Arda Pekis\\Arda Pekis: E-mail: arda.pekis@simbiosys.com}

\begin{document} 
\maketitle

\begin{abstract}
The clinical management of breast cancer depends on an accurate understanding of the tumor and its anatomical context to adjacent tissues and landmark structures. This context may be provided by semantic segmentation methods; however, previous works have been largely limited to a singular focus on the tumor alone and rarely other tissue types. In contrast, we present a method that exploits tissue-tissue interactions to accurately segment every major tissue type in the breast including: chest wall, skin, adipose tissue, fibroglandular tissue, vasculature and tumor via standard-of-care Dynamic Contrast Enhanced MRI. Comparing our method to prior state-of-the-art, we achieved a superior Dice score on tumor segmentation while maintaining competitive performance on other studied tissues across multiple institutions. Briefly, our method proceeds by localizing the tumor using 2D object detectors, then segmenting the tumor and surrounding tissues independently using two 3D U-nets, and finally integrating these results while mitigating false positives by checking for anatomically plausible tissue-tissue contacts. The object detection models were pre-trained on ImageNet and COCO, and operated on MIP (maximum intensity projection) images in the axial and sagittal planes, establishing a 3D tumor bounding box. By integrating multiple relevant peri-tumoral tissues, our work enables clinical applications in breast cancer staging, prognosis and surgical planning.
\end{abstract}

\keywords{Breast cancer, DCE MRI, Object localization, Semantic segmentation, Deep neural network, Multi-institution}

\section{Introduction}
\label{sec:intro}  % \label{} allows reference to this section

Over the past decade, Dynamic Contrast-Enhanced MRI (DCE MRI) has emerged as a standard for breast imaging. Contemporaneously, work in medical image analysis for breast cancer has focused on the segmentation of the tumor, which enables assessments of tumor dimension \cite{01Zhang_Saha_Zhu_Mazurowski_2019, 02Zhang_Luo_Chai_Arefan_Sumkin_Wu_2019, 03Hirsch_Huang_Luo_Rossi_Saccarelli_Lo_Gullo_Daimiel_Naranjo_Bitencourt_Onishi_Ko_Leithner_et_al._2022}. Few segmentation methods address the surrounding non-neoplastic tissues \cite{04Dalmış_Litjens_Holland_Setio_Mann_Karssemeijer_Gubern‐Mérida_2017, 05Wang_Morrell_Heibrun_Payne_Parker_2013, 06Zhang_Chen_Chang_Park_Kim_Chan_Chang_Chow_Luk_Kwong_et_al._2019}. Unfortunately, this limits the clinical applicability of these methods since the determination of the T stage, a key component of the American Joint Committee on Cancer (AJCC) breast cancer staging guidelines, requires not only tumor dimensions but also an evaluation of surrounding skin or chest wall involvement \cite{07Kalli_Semine_Cohen_Naber_Makim_Bahl_2018}. Additionally, tumor adjacent tissues also affect surgical planning and the growth characteristics of the cancer . The need to segment the surrounding tissues as well as the tumor highlights a need for multi-class breast MRI semantic segmentation which includes all clinically relevant tissues.

Several factors pose unique challenges for breast MRI analysis. The variability of the DCE MRI acquisition process impairs the robustness of many methods of image analysis. Scanner artifacts can include ghosting, geometric deformations, low signal-to-noise ratio, and intensity non-uniformity, which often exhibit variation over time for a single scanner, and even more dramatically between scanners \cite{08Peltonen_Mäkelä_Lehmonen_Sofiev_Salli_2020}. Other hardware variability sources include breast coil placement, and interference resulting in Moiré artifacts. Additionally, variability is introduced by the acquisition parameters such as the spatial and temporal intervals. Temporal intervals can vary from 1-30 seconds for ultra-fast DCE MRI, or up to 80-100 seconds for standard DCE MRI \cite{09Li_Newitt_Gibbs_Wilmes_Jones_Arasu_Strand_Onishi_Nguyen_Kornak_et_al._2020}. Training over a variety of image acquisition parameters, scanners and sites can mitigate unexpected performance degradation.

Additionally, breast anatomy is unique in its heterogeneous presentation. It is common for the breasts to vary greatly in size and shape between patients. Breast volume can vary by 4-fold or more \cite{10Huang_Boone_Yang_Packard_McKenney_Prionas_Lindfors_Yaffe_2011}. There is also significant variation in breast density (the ratio of glandular tissue to breast volume) \cite{11Sprague_Conant_Onega_Garcia_Beaber_Herschorn_Lehman_Tosteson_Lacson_Schnall_et_al._2016}. A major source of variability is background parenchymal enhancement (BPE), which often occurs in dense breast cases and is a major confounding factor in breast MRI interpretation \cite{12You_Kaiser_Baltzer_Krammer_Gu_Peng_Schönberg_Kaiser_2018}. These are compounded by the heterogeneous nature of breast cancer, which varies greatly in size and morphology, especially between grades such as DCIS vs invasive breast cancer. Breast cancer can also induce variation in other tissues, such as skin tagging or thickening.

\section{Prior Work and Novel Contribution}

Variability of DCE MRI can be addressed using harmonization techniques. MRI harmonization techniques were generally developed for brain MRI and adapted for use in breast MRI by prior works. It can also be addressed by using data augmentation during training to develop robust models. The use of affine or noise-based image transformations is common in medical image segmentation, however the use of elastic deformations and simulations of common MRI artifacts remains rare \cite{13Gibson_Li_Sudre_Fidon_Shakir_Wang_Eaton-Rosen_Gray_Doel_Hu_et_al._2018, 14Nalepa_Marcinkiewicz_Kawulok_2019}.

The limited availability of training data is a common concern as deep learning is best suited to settings with broad data availability. Previous works address this using self-supervised learning \cite{15Hatamizadeh_Nath_Tang_Yang_Roth_Xu_2022}, transfer learning \cite{16Liu_Xu_Zhou_Mertelmeier_Wicklein_Jerebko_Grbic_Pauly_Cai_Comaniciu_2017}, and weakly supervised learning \cite{17Chen_Hong_2022}. These approaches are rarely used in breast MRI analysis.

Semantic segmentation methods for breast MR largely target the tumor, rarely segmenting glandular or adipose tissues. Methods that use standard-of-care T1w DCE MRI are limited to segmenting tumor, adipose and glandular tissues, and typically make use of standard U-Net techniques \cite{03Hirsch_Huang_Luo_Rossi_Saccarelli_Lo_Gullo_Daimiel_Naranjo_Bitencourt_Onishi_Ko_Leithner_et_al._2022, 06Zhang_Chen_Chang_Park_Kim_Chan_Chang_Chow_Luk_Kwong_et_al._2019}. Methods using multi-parametric MRI may segment a larger variety of tissues, but clinical applicability is limited due to the non-standard MRI sequences used.

Object detection methods have previously been applied in medical imaging domains but remain rare in breast MRI analysis. They are most prevalent in lung CT analysis \cite{19Baumgartner_Jäger_Isensee_Maier-Hein_2021}. These methods are useful for reducing false positive rates, which is necessary to enable automated tumor segmentation.

\textbf{Contributions}. We present an automated method for segmenting multiple tissues on standard-of-care breast DCE MRI for clinical usage as a visual aid for surgical planning. Unlike previous works, our method delineates all clinically relevant anatomical structures of the breast simultaneously, including skin, adipose, fibroglandular, vascular, chest wall, and tumor tissues while training and evaluating on a diverse multi-institutional dataset.  This paper makes both methodological and applied contributions:

\begin{itemize}
    \item A method for utilizing 2D object detectors pre-trained on natural images for 3D object localization.
    \item A method for filtering false positive tissues in breast MR by tissue-tissue contact.
    \item We show applicability to a broader set of breast tissues than previous works.
    \item We show applicability to breast MRIs from unseen institutions with radiologist vetted ground truth.
\end{itemize}
\pagebreak
\begin{figure}
    \centering
    \includegraphics[width=0.8\textwidth]{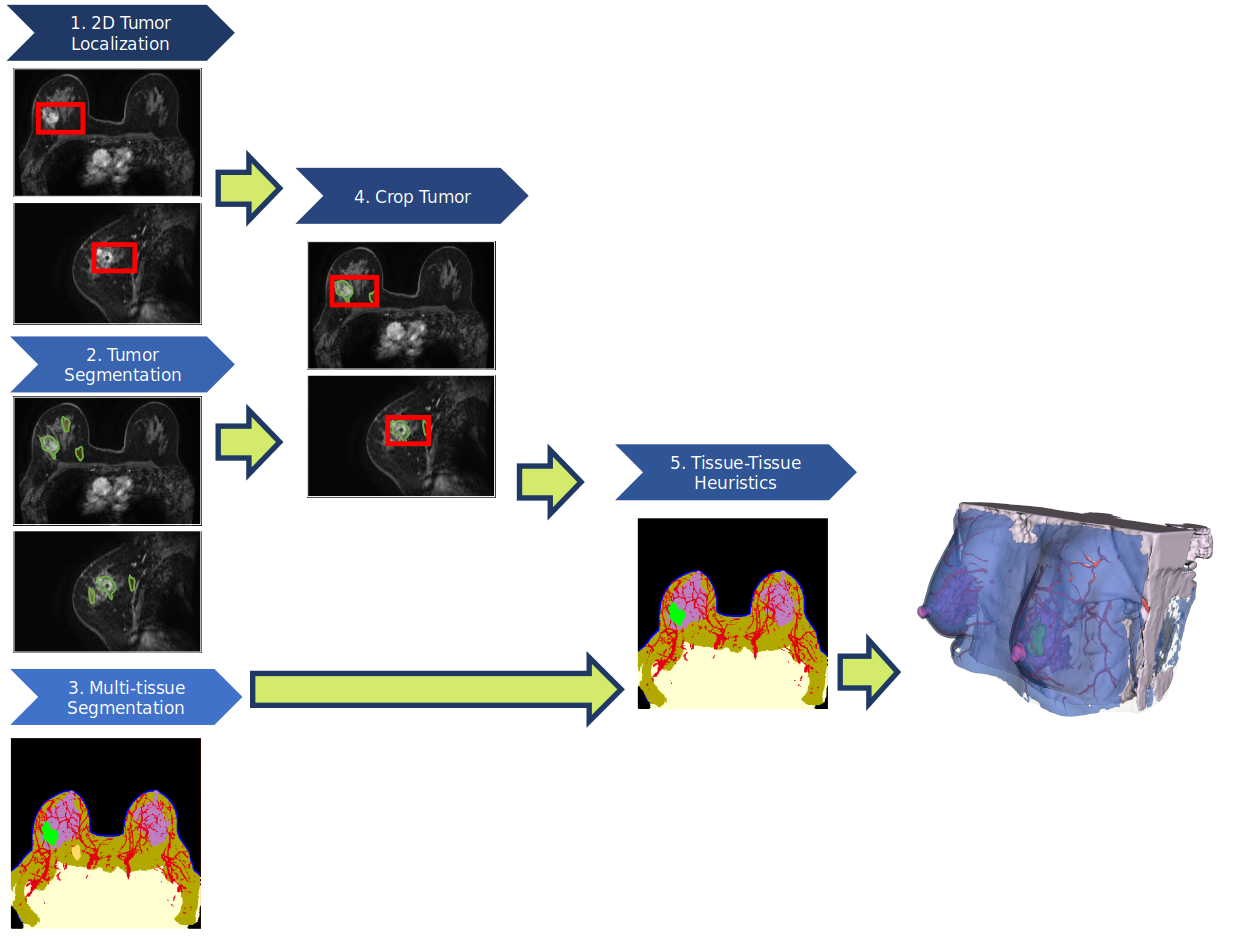}
    \caption{System diagram of our method. Modules 1-5 are described in Section \ref{sec:methods}.}
    \label{fig:system_diagram}
\end{figure}

\section{Methods}
\label{sec:methods}

Our method is a multi-phase system which uses a collection of deep learning models and post-processing steps, 

\begin{enumerate}
    \item Bi-MIP (Maximum Intensity Projection) Tumor Localization: The tumor is detected by 2D object detection networks in both the sagittal and axial MIPs.
    \item Tumor Semantic Segmentation: The tumor is segmented by a 3D semantic segmentation network.
    \item Multi-Tissue Semantic Segmentation: All tissues of interest are segmented by a 3D semantic segmentation network.
    \item Fusion of Tumor Localization and Tumor Segmentation: The 2D tumor localization bounding boxes are projected into 3D and used to crop the tumor segmentation.
    \item Tissue-Tissue Interaction Heuristics: The predictions of the tumor and multi-tissue models are merged according to heuristics developed to utilize anatomical knowledge to minimize common neural network failure modes.
\end{enumerate}

The input to the overall system is a T1-weighted (T1w) fat-suppressed DCE MRI acquired with standard-of-care parameters, and with a known disease laterality (left, right or bilateral). The output is a discrete labeling of the image with air, skin, adipose tissue, fibroglandular tissue, vasculature, tumor and chest wall. All model training was conducted on a single NVIDIA GPU with no more than 24 GB VRAM.

\subsection*{Data Preparation and Augmentation}

For each DCE MRI series, we chose three timepoints, the pre-contrast, first post-contrast (early post-contrast) and a post-contrast series at least 5 minutes after the first (late post-contrast). The images were resampled to 1 mm in all three spatial dimensions and linearly registered using phase cross correlation.

The preprocessing for the 2D localization networks consisted of applying maximum intensity projections (MIPs) through the axial and sagittal planes. The sagittal projection was limited to the half with the diseased breast. To reduce background enhancement from glandular tissue, blood vessels, and other non-tumor enhancing regions, an anisotropic median filter (window size of 10 mm) was applied along the MIP axis. Finally, the intensities are normalized to match ImageNet statistics where the timepoints selected during image processing map to the RGB channels.

When training the 3D semantic segmentation networks, we performed an extensive image augmentation routine. First, we used random cropping to a spatial extent of 128 x 128 x 64 mm$^3$ (the short axis being the Superior-Inferior axis). Then we randomly applied one or more image augmentation methods from the following list: additive gaussian noise, spatially correlated multiplicative gaussian noise, rotations, scaling, elastic deformations, and drift \cite{13Gibson_Li_Sudre_Fidon_Shakir_Wang_Eaton-Rosen_Gray_Doel_Hu_et_al._2018}. Finally, the image is normalized to the standard normal distribution.

\subsection{Bi-MIP Tumor Localization}
\label{sec:bimip}

The 2D tumor localization networks include an axial projection localization network and a sagittal projection localization network. Both networks are constructed as Shifting Windows (Swin) Transformers trained on ImageNet-1K for the backbone network of a Mask-RCNN that was trained on COCO \cite{20Chen_Wang_Pang_Cao_Xiong_Li_Sun_Feng_Liu_Xu_et_al._2019, 21He_Gkioxari_Dollar_Girshick_2017, 22Lin_Maire_Belongie_Hays_Perona_Ramanan_Dollár_Zitnick_2014, 23Liu_Lin_Cao_Hu_Wei_Zhang_Lin_Guo_2021, 24Russakovsky_Deng_Su_Krause_Satheesh_Ma_Huang_Karpathy_Khosla_Bernstein_et_al._2015, 25Vaswani_Shazeer_Parmar_Uszkoreit_Jones_Gomez_Kaiser_Polosukhin_2023}. We use the ``mm\_detection'' software library, version 2.27 to construct and train the networks \cite{20Chen_Wang_Pang_Cao_Xiong_Li_Sun_Feng_Liu_Xu_et_al._2019}.

The axial and sagittal networks were both trained for 50 epochs, with a batch size of 2 images per iteration. The target was the bounding box and mask computed from the projection of the 3D tumor segmentation onto the axial or sagittal plane. Cross entropy loss was used on the object detection and mask outputs of the networks. We used the AdamW optimizer with 1 x 10-4 learning rate, betas of 0.9 and 0.999, and weight decay of 0.05 \cite{20Chen_Wang_Pang_Cao_Xiong_Li_Sun_Feng_Liu_Xu_et_al._2019, 27Loshchilov_Hutter_2019}.

\subsection{Tumor Semantic Segmentation}
\label{sec:tumorseg}

The tumor semantic segmentation network uses a U-net architecture constructed using the ``MONAI'' software library, version 0.9 \cite{28Cardoso_Li_Brown_Ma_Kerfoot_Wang_Murrey_Myronenko_Zhao_Yang_et_al._2022}. We used the ``DynUNet'' architecture with [48, 96, 192, 384, 768] filters per block, residual skip connections, and instance normalization. Overall, the model had 51M parameters. We used the Adam optimizer, with a learning rate of 5 x 10-5 for 200 epochs. The models were trained using combo loss where the contribution of Dice component was 80\% \cite{29Taghanaki_Zheng_Zhou_Georgescu_Sharma_Xu_Comaniciu_Hamarneh_2021}. The batch size for the training procedure was 4 volumes per iteration. A positive and negative ratio of 9:1 was used for the tumor class.

\subsection{Multi-Tissue Semantic Segmentation}
\label{sec:multiseg}

We used a Residual U-net architecture, with 107M parameters \cite{06Zhang_Chen_Chang_Park_Kim_Chan_Chang_Chow_Luk_Kwong_et_al._2019}. The filter sizes were [48, 96, 192, 384] with 3 layers per block. The model was trained using the ``pytorch'' library, version 1.8 \cite{30Paszke_Gross_Massa_Lerer_Bradbury_Chanan_Killeen_Lin_Gimelshein_Antiga_et_al._2019}. Optimization was performed with stochastic gradient descent set to a learning rate of 1 x 10-4, Nesterov momentum of 0.9, and weight decay of 5 x 10-5, for 60 epochs over the training data split. The batch size was 1. Image crops were sampled uniformly. Finally, the multi-tissue segmentation model was calibrated using temperature scaling to minimize the expected calibration error. 

\subsection{Fusion of Tumor Localization and Tumor Segmentation}\
\label{sec:fusion}

The axial and sagittal box proposals form a 3D bounding box through a series of fusion and projection operations. The axial bounding box proposals are fused through a weighted average to form a single box in the diseased breast. The sagittal box proposals that overlap with the processed axial box along the anterior-posterior direction are also fused. The 3D box is then merged by averaging the dimensions from the processed boxes along the intersecting anterior-posterior plane; the remaining non-intersecting dimensions (left-right and superior-inferior) are copied over from the processed boxes. When no bounding box proposals are produced from either model, the imaging axis is used as the bounding box dimension (half the axis for the left-right plane of the diseased breast). 

\subsection{Tissue-Tissue Interaction Heuristics}
\label{sec:heuristics}

We utilize the non-tumor probabilities estimated by the multi-tissue model and the tumor probability from the tumor segmentation model. Then, we suppressed tumor connected components outside of the 3D bounding box identified above. Finally, we suppress the tumor probability by subtracting the vasculature probability and re-normalize probabilities to sum to 1. 

We analyze the contacts of various tissue components to each other to further mitigate false positives. We drop air that does not contact the edge of the image, skin which does not contact air, and tumor that does not contact glandular tissue where the contact area is less than 64 mm$^2$.Then, we used hysteresis thresholding of the tumor prediction where we include up to a 4 mm additional radius of the tumor prediction which mitigates jagged and under segmented boundaries. 

\section{Experiments}

\subsection{Development Dataset}

We sourced de-identified data retrospectively from a variety of public clinical trials and private collaborations. We collected standard-of-care T1-weighted fat-suppressed DCE MRI from the ISPY1 (32 patients) and ISPY2 (202 patients) public datasets, and seven other institutions (268 patients) \cite{31Clark_Vendt_Smith_Freymann_Kirby_Koppel_Moore_Phillips_Maffitt_Pringle_et_al._2013, 32Hylton_Gatsonis_Rosen_Lehman_Newitt_Partridge_Bernreuter_Pisano_Morris_Weatherall_et_al._2016, 34Newitt_Hylton_2016, 09Li_Newitt_Gibbs_Wilmes_Jones_Arasu_Strand_Onishi_Nguyen_Kornak_et_al._2020}. The dataset included a variety of MRI from scanners made by General Electric, Phillips, Toshiba, and Siemens. The magnetic field strength of scanners also varied between 1.5 to 3.0 Tesla.

\subsubsection{Tumor Development Data and Ground Truthing}

For the development sets of the localization and tumor segmentation models, consisted of 358 training, 44 validation, 46 test and 48 held-out patients. The hold-out set was collected from 2 institutions excluded from the training set. All tumor ground truth data were 3D annotations with 1 mm cubic resolution, identifying the extent of the solid mass of the cancer, informed by radiology reports when available. To annotate the ground truth data, a convolutional neural network was initially used to segment the tumor, and the segmentation boundaries were manually adjusted as necessary. A US-board certified radiologist advised as needed in dispute cases.

\subsubsection{Multi-Tissue Development Data and Ground Truthing}
For the multi-tissue segmentation model development dataset, there were 50 training and 15 validation patients. Annotations were performed using 3D Slicer \cite{35Fedorov_Beichel_Kalpathy-Cramer_Finet_Fillion-Robin_Pujol_Bauer_Jennings_Fennessy_Sonka_et_al._2012}. The annotation classes were air, skin, fibroglandular tissue, adipose tissue, tumor, vasculature and chest. The chest tissue class was a catch-all for the thoracic cavity, muscles, bones, and lymph nodes. All annotations had 1 mm cubic resolution. Vasculature annotations were limited to vessels above 2 mm in diameter. The computer vision tools built into 3D Slicer were used to assist in the annotation efforts. The authors performed quality control on this set. 

\subsubsection{Multi-Tissue Evaluation Dataset and Ground Truthing}
We evaluate our method on 32 complete segmentations with the tissues described previously. The patients used for the evaluation set were sourced from 3 institutions distinct from the training dataset and sent to radiologists for quality control. The radiologists reviewed 37 completed segmentations, rejecting 5 due to insufficient quality (requirement of 70\% estimated Dice score in general, 80\% for chest and adipose tissue in particular). Four exclusions were due to tumor labeling quality, and one due to vasculature. 

\subsection{Segmentation Results with Tumor Hold-out Dataset}

\begin{table}[b]
    \centering
    \caption{Tumor segmentation performance on the tumor hold-out data split. RHD is abbreviated from Robust Hausdorff Distance (95th percentile).}
    \begin{tabular}{c|c|c}
         Model Variant & Mean Dice\% (Std. err) & Mean RHD (Std. err) \\
         \hline
         Multi-Tissue & 58 (4.1) & 71 (12) \\
         + Heuristics (\S\ref{sec:heuristics}) & \textbf{64 (3.9)} & \textbf{25 (5.3)} \\
         + Local. \& Seg. (\S\ref{sec:bimip}-\ref{sec:fusion}) & \textbf{74 (3.8)} & \textbf{20 (6.6)} \\
    \end{tabular}
    \label{tab:ablation}
\end{table}

We evaluated the tissue-tissue heuristic method and tumor localization and segmentation methods for tumor segmentation performance on the hold-out data split (Table \ref{tab:ablation}). We found significant (paired t-test with $\alpha=5\%$) improvements in mean Dice score and Robust Hausdorff Distance for tumor segmentations compared to the multi-tissue model alone. Also, false positive tumor components per patient fell from 38 to 0.25 between the multi-tissue model alone and applying all methods, respectively.

\subsection{Segmentation Results with Multi-Tissue Evaluation Dataset}

\begin{table}[b]
    \centering
    \caption{Mean Dice (\%) scores (and std. errors) across tissues and model architectures were evaluated on the multi-tissue evaluation dataset (32 patients). Gland is an abbreviation of fibroglandular tissue. Bold indicates significance in a paired t-test with $\alpha=5\%$.}
    \begin{tabular}{c|c|c|c|c|c|c}
         Method & Tumor & Adipose & Gland & Vascular & Skin & Chest \\
         \hline
         nnU-Net & 56 (4.9) & 86 (1.2) & 69 (2.2) & 36 (1.4) & 71 (1.2) & 81 (2.2) \\
         Ours & \textbf{67 (4.6)} & 86 (1.1) & 67 (2.4) & 36 (1.6) & 70 (1.3) & 81 (2.1) \\
    \end{tabular}
    \label{tab:evaluation}
\end{table}

\begin{figure}[b]
    \centering
    \includegraphics[width=\textwidth]{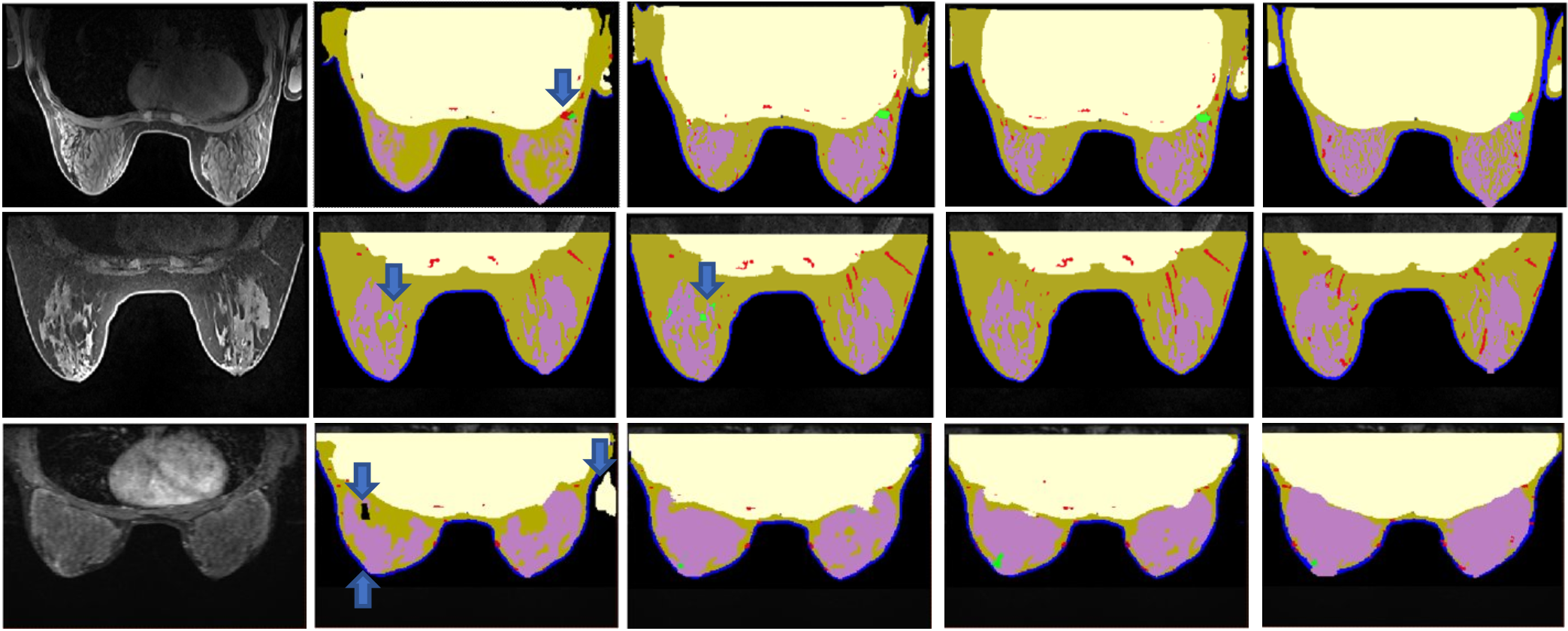}
    \caption{Examples of segmented DCE MRI. Columns represent, from left to right, MRI, nnU-Net, Residual U-Net, our approach, and reference segmentations. Tissues shown: tumor (green), gland (purple), adipose (gold), vasculature (red), chest (cream), air (black). Observe, in rows 1, 2 and 3: false positives of vasculature occur on tumor; false positives of tumor occur on glandular tissue; a false negative of tumor occurs and a false positive region of air inside the breast and chest tissue outside the body.}
    \label{fig:error-analysis}
\end{figure}

We compared our methodology to the popular nnU-Net methodology \cite{36Isensee_Jaeger_Kohl_Petersen_Maier-Hein_2021} since U-Net methods have been applied extensively to breast MRI analysis \cite{03Hirsch_Huang_Luo_Rossi_Saccarelli_Lo_Gullo_Daimiel_Naranjo_Bitencourt_Onishi_Ko_Leithner_et_al._2022, 06Zhang_Chen_Chang_Park_Kim_Chan_Chang_Chow_Luk_Kwong_et_al._2019, 36Isensee_Jaeger_Kohl_Petersen_Maier-Hein_2021}. We trained nnU-Net on a single data split, similarly to our multi-tissue model.

In evaluation on the multi-tissue evaluation dataset (Table \ref{tab:evaluation}), the presented method was on-par with nnU-Net on non-tumor tissues, while being significantly better in terms of tumor segmentation (two-tailed, paired t-test, p=0.018). The presented method was much better in terms of tumor false positive connected component counts, with an average of 0.48 per patient, compared to 3.8 for nnU-Net. We qualitatively compare the inference results in Figure \ref{fig:error-analysis}.

\pagebreak
\section{Concluding Remarks}

Previous models were limited in robustly segmenting multiple breast tissues using standard-of-care DCE MRI across sites.  Here, we presented the first 3D multi-tissue deep learning model for breast MRI that encompasses clinically relevant non-neoplastic tissues including: adipose tissue, glandular tissue, vasculature, skin, and chest wall, while also achieving excellent tumor segmentation performance. Our work is a step towards the broader vision of panoptic segmentation in medical imaging and opens new avenues for incorporating 3D visualization in clinical applications for comprehensive tumor staging, surgical planning, biomedical imaging research, investigations into tumor biology, and patient education.

\acknowledgments % equivalent to \section*{ACKNOWLEDGMENTS}       
 
Thanks to the data annotators for providing the segmentations needed for development of the model. Thanks to Anant Madabhushi for helpful discussions regarding the presentation of the work.

% References
\bibliography{report} % bibliography data in report.bib
\bibliographystyle{spiebib} % makes bibtex use spiebib.bst

\end{document}